\documentclass[prl,showpacs,superscriptaddress,twocolumn]{revtex4}

\usepackage{amsmath,bm}
\usepackage{graphicx}
\usepackage{epsfig}	

\begin{document}

\newcommand{\vk}{{\vec k}}
\newcommand{\vK}{{\vec K}}
\newcommand{\vb}{{\vec b}}
\newcommand{{\vp}}{{\vec p}}
\newcommand{{\vq}}{{\vec q}}
\newcommand{\vQ}{{\vec Q}}
\newcommand{\vx}{{\vec x}}
\newcommand{\beq}{\begin{equation}}
\newcommand{\eeq}{\end{equation}}
\newcommand{\half}{{\textstyle \frac{1}{2}}}
\newcommand{\gton}{\stackrel{>}{\sim}}
\newcommand{\lton}{\mathrel{\lower.9ex \hbox{$\stackrel{\displaystyle<}{\sim}$}}}
\newcommand{\ee}{\end{equation}}
\newcommand{\ben}{\begin{enumerate}}
\newcommand{\een}{\end{enumerate}}
\newcommand{\bit}{\begin{itemize}}
\newcommand{\eit}{\end{itemize}}
\newcommand{\bc}{\begin{center}}
\newcommand{\ec}{\end{center}}
\newcommand{\bea}{\begin{eqnarray}}
\newcommand{\eea}{\end{eqnarray}}

\newcommand{\beqar}{\begin{eqnarray}}
\newcommand{\eeqar}[1]{\label{#1} \end{eqnarray}}
\newcommand{\pleft}{\stackrel{\leftarrow}{\partial}}
\newcommand{\pright}{\stackrel{\rightarrow}{\partial}}

\newcommand{\eq}[1]{Eq.~(\ref{#1})}
\newcommand{\fig}[1]{Fig.~\ref{#1}}
\newcommand{\eff}{ef\!f}
\newcommand{\alphas}{\alpha_s}

\renewcommand{\topfraction}{0.85}
\renewcommand{\textfraction}{0.1}
\renewcommand{\floatpagefraction}{0.75}

\title{ Transverse Momentum Balance and Angular Distribution of $b\bar{b}$ Dijets in Pb+Pb collisions}

\date{\today  \hspace{1ex}}
\author{Wei Dai}
\affiliation{School of Mathematics and Physics, China University
of Geosciences, Wuhan 430074, China}

\author{Sa Wang}
\affiliation{Key Laboratory of Quark \& Lepton Physics (MOE) and Institute of Particle Physics,
 Central China Normal University, Wuhan 430079, China}

\author{Shan-Liang Zhang}
\affiliation{Key Laboratory of Quark \& Lepton Physics (MOE) and Institute of Particle Physics,
 Central China Normal University, Wuhan 430079, China}

\author{Ben-Wei Zhang\footnote{bwzhang@mail.ccnu.edu.cn}}
\affiliation{Key Laboratory of Quark \& Lepton Physics (MOE) and Institute of Particle Physics,
 Central China Normal University, Wuhan 430079, China}
\affiliation{Guangdong Provincial Key Laboratory of Nuclear Science, Institute of Quantum Matter, South China Normal University, Guangzhou 510006, China}

\author{Enke Wang}
\affiliation{Guangdong Provincial Key Laboratory of Nuclear Science, Institute of Quantum Matter, South China Normal University, Guangzhou 510006, China}
\affiliation{Key Laboratory of Quark \& Lepton Physics (MOE) and Institute of Particle Physics,
 Central China Normal University, Wuhan 430079, China}

\begin{abstract}
The productions of inclusive b-jet and $b\bar{b}$ dijets in Pb+Pb collisions have been investigated by considering  the heavy quark and the light quark in-medium evolution simultaneously. The initial hard processes of inclusive b-jet and $b\bar{b}$ dijets productions are described by a next-to-leading order (NLO) plus parton shower Monte Carlo (MC) event generator SHERPA which can be well matched with the experimental data in p+p collisions. The framework combines the Langevin transport model to describe the evolution of bottom quark also its collisional energy loss and the higher-twist description to consider the radiative energy loss of both bottom and light quarks. We compare the theoretical simulation of inclusive jet and inclusive b-jet $R_{\rm AA}$ in Pb+Pb collisions at $\sqrt{s_{\rm NN}}=2.76$~TeV with the experimental data, and then present the theoretical simulation of the momentum balance of the $b\bar{b}$ dijet in Pb+Pb collisions at $5.02$~TeV with the recent CMS data for the first time. A similar trend as that in inclusive dijets has been observed in $b\bar{b}$ dijets, the production distribution is shifted to smaller $x_J$ due to the jet quenching effect.  At last, the prediction of the normalized azimuthal angle distribution of the $b\bar{b}$ dijet in Pb+Pb collisions at $5.02$ TeV has been reported. The medium induced energy loss effect of the $b\bar{b}$ dijets will overall suppress its production, but the same side ($\Delta \phi \to 0$ region) suffers more energy loss than away side  ($\Delta \phi \to \pi$ region), therefore lead to the suppression on the same side and the enhancement on the away side in the normalized azimuthal angle distribution in A+A collisions.
\end{abstract}


\maketitle

\section{Introduction}
To probe the properties of the quark-gluon plasma (QGP) in heavy-ion collisions(HIC), the medium modification of the energetic partons which produced from the initial hard scattering and later propagate through the fireball has long been investigated. It is referred as the jet quenching phenomenon, enormous effort has been devoted to explore such effect on different observables: from high $p_{\rm T}$ hadron production suppression to inclusive reconstructed full jets and even less inclusive jets such as dijets,  tagged jets etc, as well as a variety of jet substructures~\cite{Wang:1991xy,Gyulassy:2003mc, Qin:2015srf, Vitev:2008rz, Vitev:2009rd, Qin:2010mn, CasalderreySolana:2010eh,Young:2011qx,He:2011pd,ColemanSmith:2012vr,Zapp:2012ak,Ma:2013pha, Senzel:2013dta, Casalderrey-Solana:2014bpa,Milhano:2015mng,Chang:2016gjp,Majumder:2014gda, Chen:2016cof, Chen:2016vem, Chien:2016led, Apolinario:2017qay,Connors:2017ptx,Dai:2012am,Wang:2013cia,Zhang:2018urd,Neufeld:2010fj,Huang:2013vaa,Huang:2015mva,Cao:2015hia,Chen:2019gqo,Yan:2020zrz,Chen:2020pfa}.  Among them, the transverse momentum imbalance of the inclusive dijets is one of the early and fundamental observables. The jet quenching effect give a net imbalance to the $p_{\rm T}$ distributions of the back-to-back jets that can exceed the imbalance brought by the QCD correction. This additional imbalance is due to the energy losses that these two jets suffered when they propagated through the QGP medium respectively. The flavor dependence of the parton that initiate the jets is quite essential to the underlying dynamics of the jet in-medium evolution and energy loss. In the inclusive jets and inclusive dijets events, it is not easy for the experimentalists to determine such dependence, however the pairs production of the heavy quark $b\bar{b}$ jets that are back-to-back in azimuth provide a golden chance to isolate the type of the parton that initiate the jets. It is simply because in this scenario, the produced jets can be restricted to be $b(\bar{b})$ quark initiated. It is of great interest to compare the possible different medium modification effects of the $b\bar{b}$ dijets with inclusive dijets since, in inclusive dijets events, jets are predominately initiated by light quark and gluon.

Both ATLAS~\cite{Aaboud:2016jed} and CMS~\cite{Sirunyan:2018jju} collaborations reported their measurements of the $b\bar{b}$ dijets production in p+p collision at the LHC comparing with several Next-to-Leading Order (NLO) and Leading Order (LO) QCD Monte Carlo simulations, but the theoretical prediction of the medium modification of $b\bar{b}$ dijets in Pb+Pb collisions is required to confront with the available CMS data~\cite{Sirunyan:2018jju}.

There are multi mechanisms for $b\bar{b}$ dijets productions in p+p collision, one need not only enhance the production of the desired scenario, and also precisely calculate the productions from other mechanisms so that we can quantitively compare with the experimental data. Another challenge in the study of their productions in A+A collisions is that the simultaneous description is required to consider the in-medium evolution of the heavy and light quarks. So the manuscript is organized as follows: we first introduce the production mechanisms of the $b\bar{b}$ pair and also the p+p baseline of $b\bar{b}$ dijets: both the simulation setup and the event selection. Then, the framework of simultaneous description of heavy and light quark in-medium evolution is introduced. Afterwards, we will present the theoretical results of the transverse momentum imbalance distribution of the $b\bar{b}$ dijets as a function of $p_{\rm T}$ to compare with the CMS data in A+A collisions. We will also analyze the difference between the $b\bar{b}$ dijets and the inclusive dijets with respect to the transverse momentum imbalance. At last, we will report the prediction of the normalized azimuthal angle distribution of the $b\bar{b}$ dijet in Pb+Pb collisions at $5.02$ TeV to compared with its p+p reference.

\section{ Theoretical framework}

By definition, a $b(\bar{b})$ jet is a full jet with at least one $b$ or $\bar{b}$ quark inside the jet cone with the jet radius parameter $R$.  The production of a $b\bar{b}$ pair could be categorized into three production mechanisms which can be used to understand the $b\bar{b}$ dijet system~\cite{Norrbin:2000zc,Combridge:1978kx,Nason:1987xz,Beenakker:1990maa}. The flavor creation (FCR) describes the production of both $b$ and $\bar{b}$ jets that originated from the $b$ and $\bar{b}$ quarks produced back-to-back in azimuthal from the initial hard scattering, therefore these jets are supposed to be the hardest in the event. This is the process that is perfect to isolate the type the parton (b quark) that initiates a jet. The gluon splitting (GSP) mechanism allows a pair of $b(\bar{b})$ jets initiating from the $b$ and $\bar{b}$ quarks created from gluon splitting process $g \to b \bar{b} $. This pair of jets is expected to propagate in the same direction. However, the flavor excitation (FEX) mechanism is more complex and the produced $b(\bar{b})$ jet pair is more likely neither back-to-back nor on the same side. The NLO calculation without further kinetic constraint indicates that there are large contribution fractions from all three mechanisms in the investigated $p_{\rm T}$ region~\cite{Banfi:2007gu}. Therefore, a suitable kinetic constraint should be imposed to select the desired mechanism. For instance, to look at pairs of $b(\bar{b})$ jets that are constrained back-to-back in azimuth experimentally, one may largely reduce the contribution from GSP and mainly focus on the FCR process by imposing energetic $p_{\rm T}$ triggers and also intersection angle restriction between the two $b\bar{b}$ jets. This configuration are essential to provide a less ambiguous observable.

It has been demonstrated in both ATLAS~\cite{Aaboud:2016jed} and CMS~\cite{Sirunyan:2018jju} reports that the NLO effects are essential for the modeling of such observable since NLO QCD calculation with POWHEG give a better description than PYTHIA 6 alone. It is also noted that the configurations are slightly different in these two experimental publications.

\begin{figure}[!t]
\begin{center}
\vspace*{-0.2in}
\hspace*{-.1in}
\includegraphics[width=3.4in,height=2.8in,angle=0]{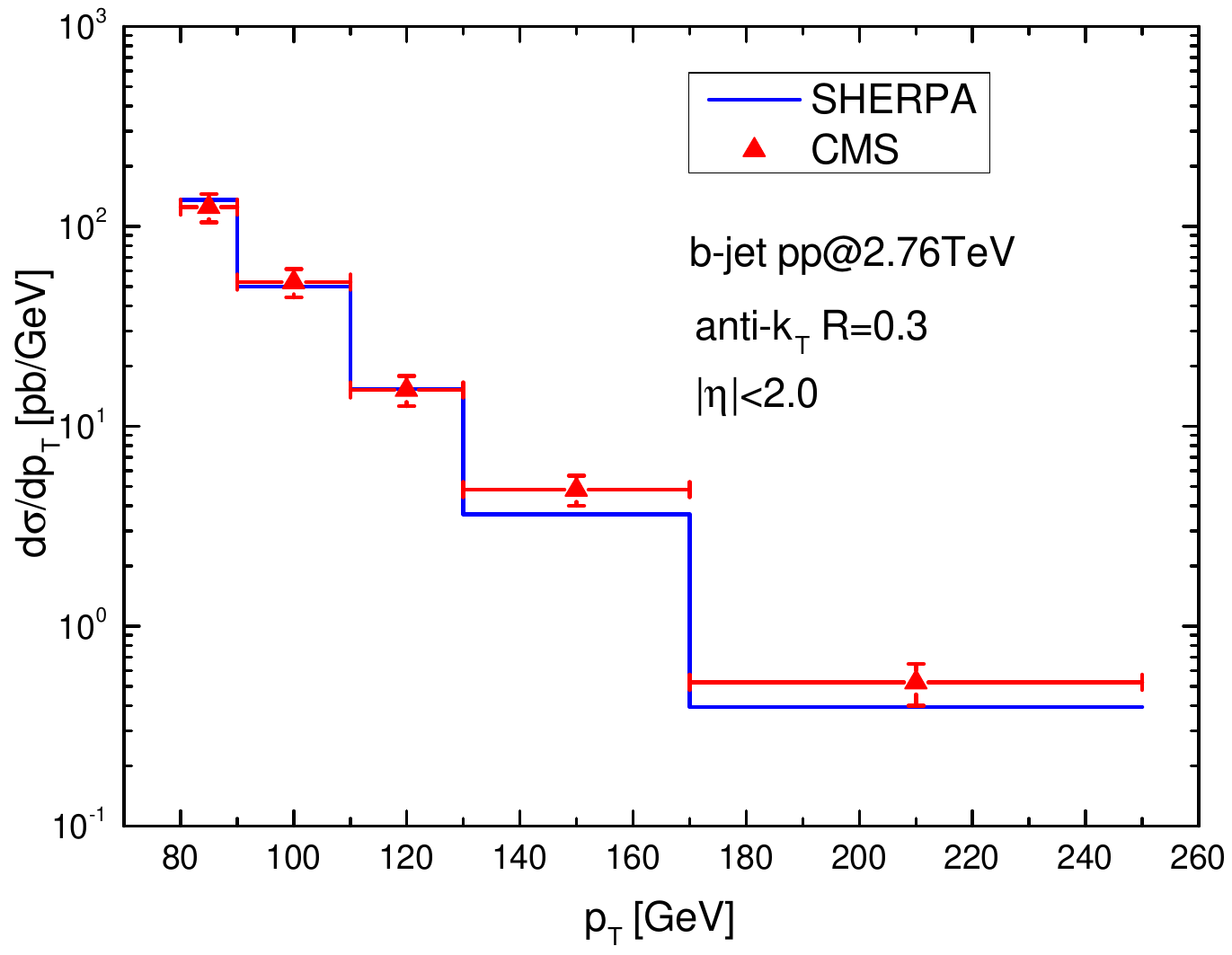}
\includegraphics[width=3.4in,height=2.8in,angle=0]{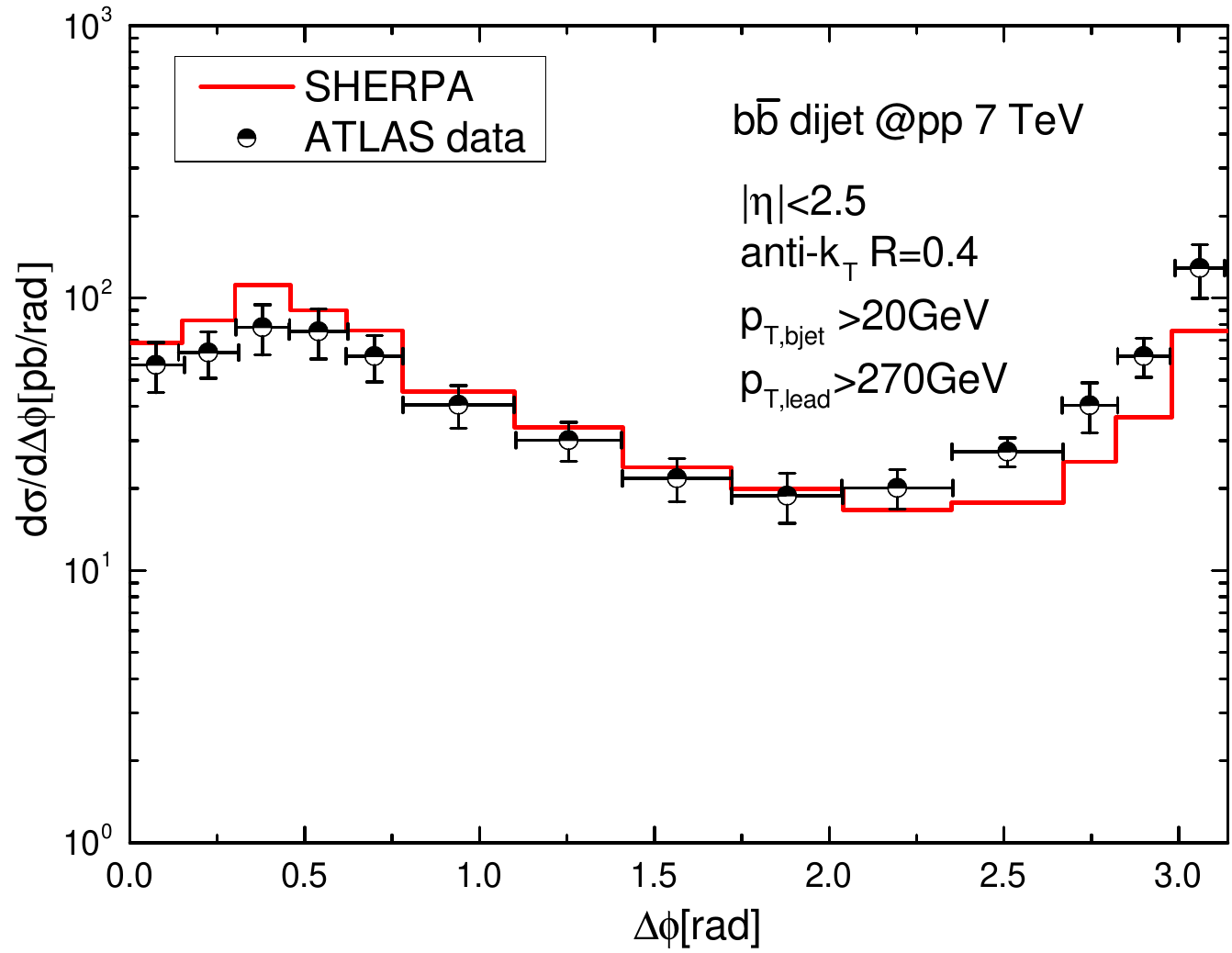}
\vspace*{-.1in}
\caption{Upper: NLO+PS result of the b-jet production in p+p collisions at $\sqrt{s_{\rm NN}}=2.76$~TeV calculated in SHERPA(in the vertical line) is compared with CMS data (in red points with error bars)~\cite{Jung:2014hja}. Bottom: NLO+PS differential cross section of the $b\bar{b}$ dijets production in p+p collision at $\sqrt{s_{\rm NN}}=7$~TeV as a function of the azimuthal angle between the two b-jets $\Delta \phi$ calculated by SHERPA is confronted with ATLAS data~\cite{Aaboud:2016jed}.}

\label{fig:illustpp}
\end{center}
\end{figure}

In our simulation, we employ the NLO+parton shower (PS) event generator SHERPA 2.2.4~\cite{Gleisberg:2008ta} to provide the inclusive jets and also inclusive dijets events in p+p collision. We select the single b-jets and $b\bar{b}$ dijets events from these inclusive events as our p+p baseline. In the SHERPA generator, we set the tree-level matrix elements are calculated by Amegic~\cite{Krauss:2001iv} and Comix~\cite{Gleisberg:2008fv} while one-loop matrix elements are calculated by BlackHat~\cite{Schumann:2007mg}. The parton shower is implemented based on the Catani-Seymour subtraction method~\cite{Nahrgang:2013saa}. The matching of NLO QCD matrix elements with the parton shower is using the MC@NLO method~\cite{Frixione:2002ik}. The NLO PDF sets in NNPDF3.0~\cite{Ball:2014uwa} with 5-flavour have been chosen in our simulation. FASTJET~\cite{Cacciari:2011ma} with anti-$k_{\rm T}$ algorithm is used for event selection and final state jet reconstruction. The jets are defined in a cone with the jet radius parameter $\Delta R=\sqrt{(\Delta \phi)^2+(\Delta \eta)^2}$, where $\phi$ and $\eta$ are the azimuthal angle and the rapidity of particles. Corresponding configuration has been set up to be in line with each p+p measurement in CMS~\cite{Jung:2014hja} and ATLAS~\cite{Aaboud:2016jed}.

In order to test the setups to generate b-jet events in SHERPA, in the upper plots of Fig.~\ref{fig:illustpp}, we present the confrontation of the theoretical simulation results of the inclusive b-jets production in p+p collision at $\sqrt{s_{\rm NN}}=2.76$~TeV with the CMS data. A good agreement has been found. We further select the $b\bar{b}$ dijets events from the SHERPA generated inclusive dijets events which can naturally include all three production mechanisms of the $b\bar{b}$ dijets, and then reproduce the differential cross section of the $b\bar{b}$ dijets as a function of the azimuthal angle between the two b-jets. A good description of the ATLAS data is shown in the bottom plots of Fig.~\ref{fig:illustpp}. It demonstrate that the azimuthal angle distribution of the two b-jets with a certain $p_{\rm T}$ trigger of the $b\bar{b}$ events. It is noted that there is a same side peak in the small azimuthal angle region which is unusual for the double inclusive observables we have investigated earlier. The fairly good description of the experimental data indicate that the higher order correction and its matched PS provided by SHERPA are essential for a solid p+p baseline. We also find that this kind double peak distribution is sensitive to the imposed kinetic cut. It requires the leading b-jet $p_{\rm T}>270$~GeV, and the lower threshold of the $p_{\rm T}$ cut of the b-jet is relatively small $p_{\rm T, b_{\rm jet}}>20$~GeV in the ATLAS publication. With the increasing of the minimum requirement of the b-jet $p_{\rm T}$ to $40$~GeV, we find the same side peak begins to vanish. From this observation, we can conclude that the double peak structure is mainly due to the contribution from the GSP process, and if one need to focus on the FCR process to have more proportion of the b-jets case to be b quark initiated, a relatively higher $p_{\rm T}$ cut of the lower threshold of the b-jet $p_{\rm T, b_{\rm jet}}$ will help. With this basic knowledge of the $b\bar{b}$ dijets production and the good performance of the SHERPA simulation, we have a very good baseline to investigate the in-medium modification of the $b\bar{b}$ dijets.


At the moment, the exact mechanism of the in-medium interaction between the heavy quarks and the QCD medium is still an open question which has been extensively investigated in both perturbative and non perturbative approaches~\cite{Moore:2004tg,Rapp:2018qla}. Transport models such as Langevin and Boltzmann approaches incorporated with the evolution profile of the bulk medium are been employed for the description of the heavy quark in-medium evolution ~\cite{Scardina:2017ipo,Djordjevic:2013xoa,Moore:2004tg, Cao:2013ita, Cao:2017hhk, Nahrgang:2013saa,Beraudo:2015wsd,Kang:2016ofv,He:2011zx,Lang:2012cx,Sharma:2009hn}. In a framework that heavy and light quark in-medium evolution and energy loss can be taken into account simultaneously, we employe a modified Langevin transport equation with an additional radiation term to include the radiative energy loss to describe the transport and the energy loss (elastic and inelastic) of heavy quark in the hot and dense medium~\cite{Cao:2013ita,Zhou:2016vwq,Wang:2019xey}:

\begin{eqnarray}
&\vec{x}(t+\Delta t)=\vec{x}(t)+\frac{\vec{p}(t)}{E}\Delta t \\
&\vec{p}(t+\Delta t)=\vec{p}(t)-\Gamma\vec{p} \Delta t+\vec{\xi}(t)-\vec{p}_g
\label{eq:lagevin}
\end{eqnarray}

where $\Delta t$ is the evolution time step defined in the simulation, $\Gamma$ is the drag coefficient which can control the strength of the elastic energy loss, $\vec{\xi}(t)$ is the stochastic term representing the random kicks by quasi-particles in such thermal medium, it obeys $\left \langle \xi^i(t)\xi^j(t') \right \rangle =\kappa \delta^{ij}\delta(t-t')$, where  $\kappa$ is the diffusion coefficient. The classic fluctuation-dissipation relation~\cite{Kubo} between $\Gamma$ and $\kappa$ has been employed:

\begin{eqnarray}
\kappa=2\Gamma ET=\frac{2T^2}{D_s}
\label{eq:fdt}
\end{eqnarray}

where $D_s$ is the spatial diffusion coefficient. Over the years, $D_s$ is used as the parameter to represent the strength of the elastic interaction between the heavy quark and the thermal medium, it has been calculated by many theoretical models~\cite{Gossiaux:2009qf,Plumari:2011mk,Hosotani:2007kn,vanHees:2004gq}. We note that Lattice calculation predicted a range of the value of $D_s$ as $2\pi TD_s\sim 3.7-7.0$~\citep{Francis:2015daa,Rapp:2018qla}. The inclusion of the last term $\vec{p}_g$ in the momentum update equation is an effective treatment and it assume the radiative energy loss of the heavy quark is carried away by the radiative gluon. The calculation of such term is based on the Higher-Twist scheme~\cite{Guo:2000nz,Zhang:2003wk,Zhang:2004qm, Majumder:2009ge}, it can provide the radiative gluon spectrum:

\begin{eqnarray}
\frac{dN}{dxdk^{2}_{\perp}dt}=\frac{2\alpha_{s}C_sP(x)\hat{q}}{\pi k^{4}_{\perp}}\sin^2(\frac{t-t_i}{2\tau_f})(\frac{k^2_{\perp}}{k^2_{\perp}+x^2M^2})^4
\label{eq:dndxdk}
\end{eqnarray}

where $x$ and $k_\perp$ are the energy fraction and the transverse momentum of the radiated gluon, $M$ is the mass of the parent parton. In addition, $C_s$ is the quadratic Casimir in color representation, and $P(x)$ is the splitting function in vacuum~\cite{Wang:2009qb}, $\tau_f=2Ex(1-x)/(k^2_\perp+x^2M^2)$ is the gluon formation time. $\hat{q}$ is the jet transport parameter proportional to the local parton density in medium when the jet probed. The time-space evolution of the QCD medium thus can be taken into account by altering the value of $\hat{q}$ relative to its initial value $\hat{q}_0$ in the most center of the overlap region at the initial time when QGP formed~\cite{Chen:2010te}. Therefore, $\hat{q}_0$ is the other parameter left controlling the strength of the bremsstrahlung jet-medium interaction.

In the simulation, the particles listed in the p+p events with the full vacuum parton-shower produced by SHERPA with their initial positions sampled from Glauber Model are served as the input of the in-medium evolution. The heavy quark will evolve in the QCD medium with the modified Langevin formalism described above in a fixed evolving time step when the position and momentum of the light quark update simultaneously. A Possion probability distribution is implemented to compare with uniform random number in order to determine whether the radiative energy loss occurs or not in a given Langevin evolution time step for both heavy and light quark, expressed as

\begin{eqnarray}
P_{rad}(t,\Delta t)=1-e^{-\left\langle N(t,\Delta t)\right\rangle}
\label{eq:poss}
\end{eqnarray}
where the $\left\langle N(t,\Delta t)\right\rangle$ is the averaged radiative gluon number in the fixed update time step $\Delta t$ at certain evolution time $t$ and can be derived by integrating Eq.\ref{eq:dndxdk}. If radiation occurs, the number of the radiated gluon then would be sampled by the above distribution. The $x$ and $k_\perp$ could be sampled according to the radiative gluon spectrum in Eq.\ref{eq:dndxdk} to give the momentum of the radiated gluon, therefore the $\vec{p}_g$ term of the Eq.\ref{eq:lagevin} in each time interval is determined. One should note that the four-momentum of the heavy quark will firstly be boosted into the local rest frame and then update according to Eq.\ref{eq:lagevin} and boosted back to the laboratory frame every evolution time step so that it can update its position there.

The smooth iEBE-VISHNU hydro model~\cite{Shen:2014vra} has been used to provide the evolution information of this hot and dense medium. During the in-medium simulation, each parton propagates in the expanding medium until the probed temperature of the local medium is under $T_c=165$~MeV.  In this manuscript, we directly set the free parameter $\hat{q}_0=1.2$~GeV$^2$/fm which is the best value taken from the global extraction of the identified hadron production in Pb+Pb collisions at $2.76$~TeV in our previous work~\cite{Ma:2018swx} since, the property of the QGP medium is also described by smooth iEBE-VISHNU hydro model there.

It is noted that the treatment of including radiative energy loss in the Langevin equation in Eq.~(2) is an effective approach to simulate heavy quark in-medium evolution since it is not easy for one to include radiative energy loss without breaking the fluctuation-dissipation relation. A lower energy cut to the radiative gluon is imposed to make sure the heavy quark can reach thermal equilibrium at low $p_{\rm T}$ regime since it can naturally be dominated by elastic energy loss at such regime~\cite{Cao:2013ita,Cao:2018ews}. There are actually two free parameters in this framework, the $\kappa$ controlling the elastic energy loss of the heavy quark, the $\hat{q}_{0}$ controlling the strength of medium induced radiative energy loss of both light and heavy quark. Besides, we neglect the collisional energy loss of light quark and gluon, due to their small contributions to the total energy loss of the light quark at larger $p_{\rm T}$~\cite{Burke:2013yra}. Recently this framework has been extended to the study of medium modifications of radial distributions of $D^0$ meson inside jets in Pb+Pb collisions relative to that in p+p at the LHC, and a decent agreement between the model calculations and the experiment measurement has been observed~\cite{Wang:2019xey} .
\section{Results and discussion}
\begin{figure}[!t]
\begin{center}
\vspace*{-0.2in}
\hspace*{-.1in}
\includegraphics[width=3.4in,height=3.in,angle=0]{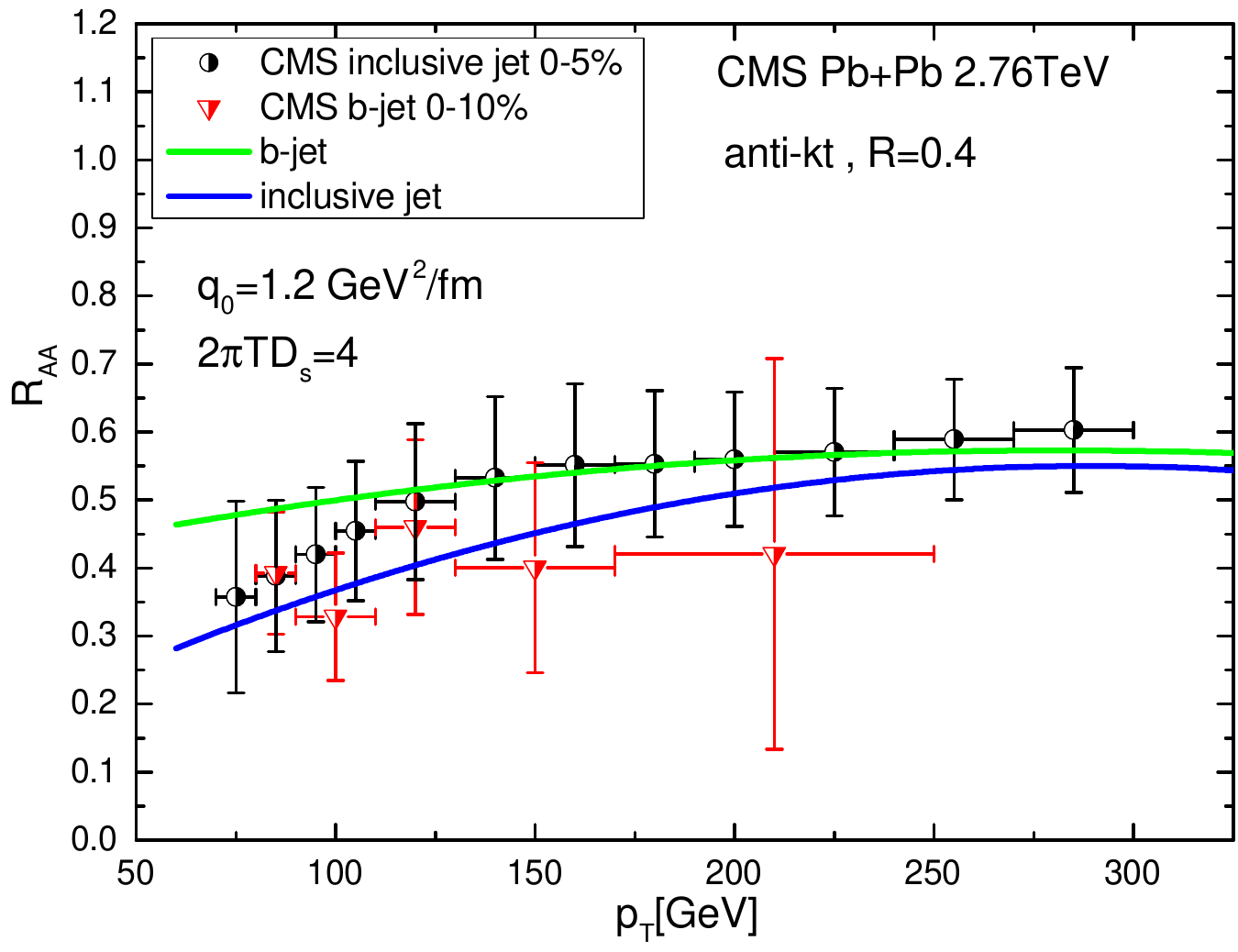}

\vspace*{-.2in}
\caption{$R_{\rm AA}$ of the inclusive jets and the inclusive b-jets as functions of $p_T$ are theoretically simulated in the same framework are compared with the CMS data ~\cite{Khachatryan:2016jfl,Jung:2014hja} respectively. }
\label{fig:model}
\end{center}
\end{figure}

Having the SHERPA generated p+p events evolved in the above framework, we obtain the final state partons which include produced partons, jet shower, radiated gluon after their in-medium modification. Implementing jet reconstruction as well as the jet selection through FASTJET~\cite{Cacciari:2011ma} on these final state partons, we could derive the jets productions in Pb+Pb collisions at the LHC. Comparing the productions with their p+p counterparts, we can study the medium modification of jet observables. To test the validation of such framework and also to verify its performance, we firstly calculated the nuclear modification factor with respect to the jet $p_{\rm T}$ of both inclusive jets and inclusive b-jets in the Pb+Pb collisions at the LHC $\sqrt{s}=2.76$~TeV to compare with the experimental data~\cite{Khachatryan:2016jfl,Jung:2014hja}. We find, with the spatial diffusion factor $D_s$ extracted from Lattice which satisfies $2\pi TD_s=4.0$ and jet transport parameter extracted from hadron suppression study which is $\hat{q}_0=1.2$~GeV$^2$/fm mentioned in the last section, our simultaneous simulation for both inclusive jets and inclusive b-jets $R_{\rm AA}$ can describe the CMS data~\cite{Jung:2014hja} fairly well within the margin of error, only the simulation for inclusive b-jet $R_{\rm AA}$ slightly overestimates the CMS data shown in Fig.~\ref{fig:model} which is similar with the calculation presented in Ref.~\cite{Huang:2013vaa,Huang:2015mva}.  With the chosen parameters mentioned above, our prediction indicate that, at lower $p_{\rm T}$ region, the heavy quark jets suffer less energy loss then the light ones, but the mass effect of the jet quenching tend to disappear when it reaches to the higher $p_{\rm T}$ regime where $R_{\rm AA}$ of the inclusive b-jets coincide with that of the inclusive jets. The establishment of such evolution framework allows us to investigate the medium modification of the $b\bar{b}$ dijets production in A+A collisions.

\begin{figure}[!t]
\begin{center}
\vspace*{-0.2in}
\hspace*{-.1in}
\includegraphics[width=3.4in,height=2.6in,angle=0]{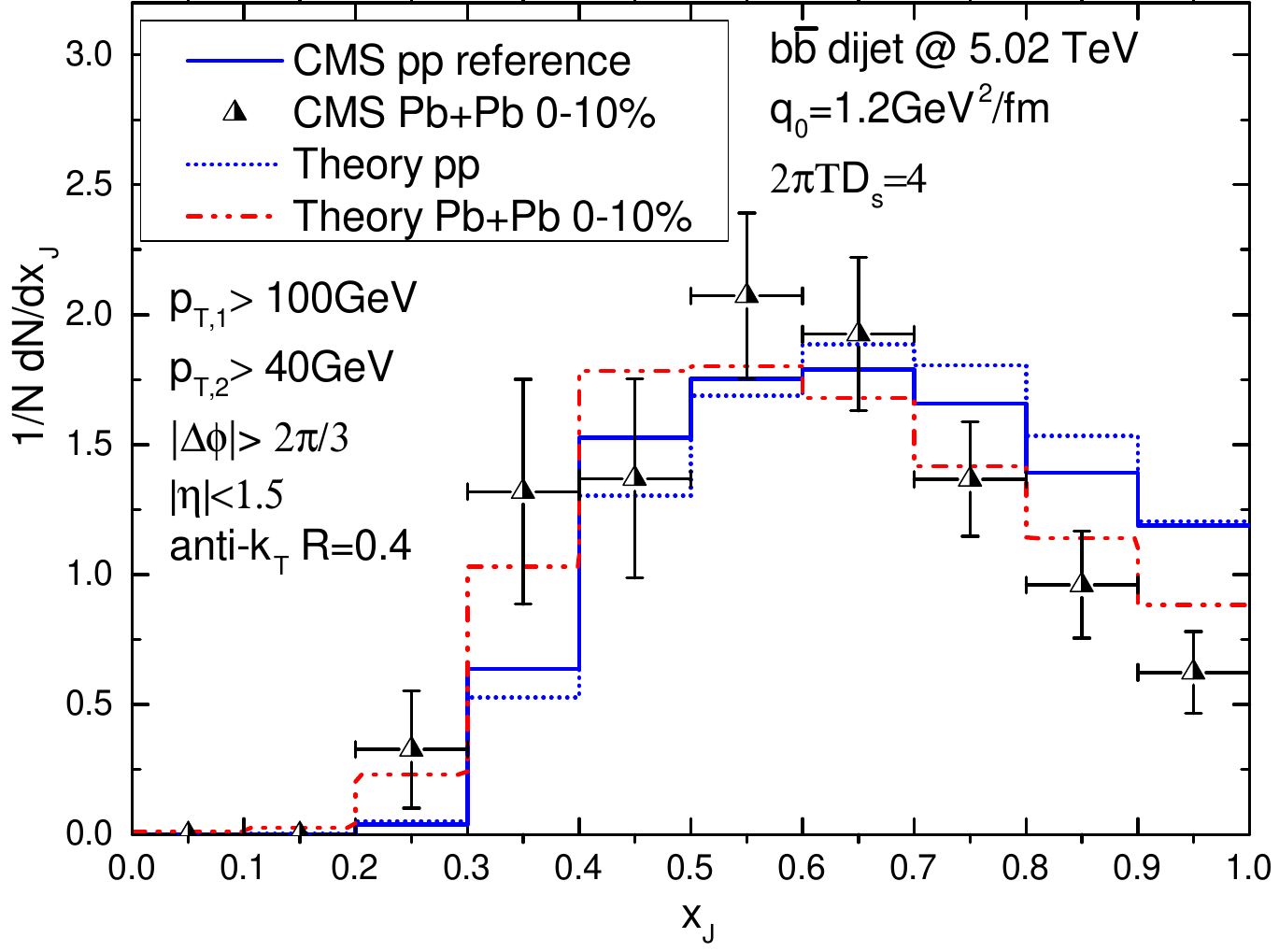}
\includegraphics[width=3.4in,height=2.6in,angle=0]{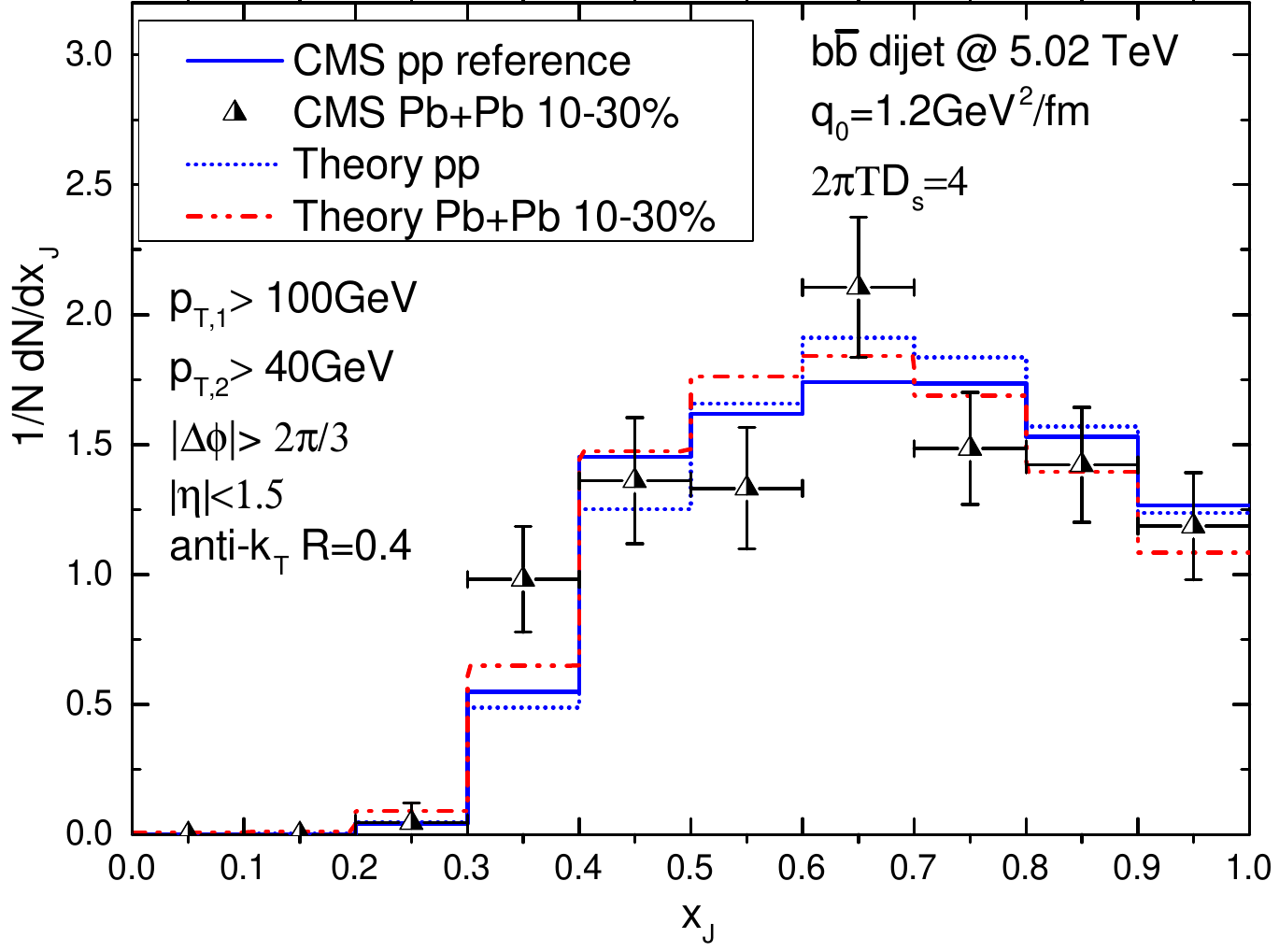}
\vspace*{-.2in}
\caption{Upper: Calculated normalized $x_{\rm J}$ distribution of $b\bar{b}$ dijets in p+p and $0-10\%$ Pb+Pb collisions at $5.02$~TeV compared with the smeared p+p baseline and experimental data in A+A collisions. Bottom: Calculated normalized $x_J$ distribution of $b\bar{b}$ dijets in p+p and $10-30\%$ Pb+Pb collisions at $5.02$~TeV compared with the smeared p+p baseline and experimental data in A+A collisions.  }
\label{fig:xj}
\end{center}
\end{figure}

The momentum balance of the $b\bar{b}$ dijets defined as the $p_{\rm T}$ ratio of the sub-leading jet to the leading jet: $x_J=p_{\rm T,2}/p_{\rm T,1}$. These two leading jets are required to be b-jets. We demonstrate the simulation results of the normalized distributions of $x_{\rm J}$ in p+p and Pb+Pb collisions for $b\bar{b}$ dijets to compare with experimental data. As the same selection with the CMS experiment, we set the minimum $p_{\rm T}$ cut of the leading and the sub-leading jets to be $100$~GeV and $40$~GeV respectively. Further, the selection of $\vert \Delta \phi \vert > 2\pi/3$ has also been applied to require the opening angle of the two jets are back-to-back in azimuthal both in p+p and A+A collisions. A same smearing treatment suggested by CMS has been performed to confront with the CMS data shown in Fig.~\ref{fig:xj}. Noted that the p+p reference in experiment is obtained from each jet $p_{\rm T}$ data smeared by resolution parametrization at given centrality, we find our results are in consistent with both p+p and Pb+Pb experimental measurement at $5.02$~TeV. The energy loss effect will suppress the distribution at larger $x_{\rm J}$ region and enhance it at lower $x_{\rm J}$. It therefore lead to a lower shift of the overall $x_{\rm J}$ distribution. We note that the shift of the A+A $x_{\rm J}$ distribution relative to p+p reference is quite visible in central collisions shown in the left plots. Much smaller shift is observed at $10-30\%$ Pb+Pb collision shown in the bottom panel of the Fig.~\ref{fig:xj}, suggesting smaller energy loss suffered in more peripheral collisions which is in consistent with the case in dijets~\cite{Sirunyan:2018jju}.

\begin{figure}[!t]
\begin{center}
\vspace*{-0.2in}
\hspace*{-.1in}
\includegraphics[width=3.4in,height=2.8in,angle=0]{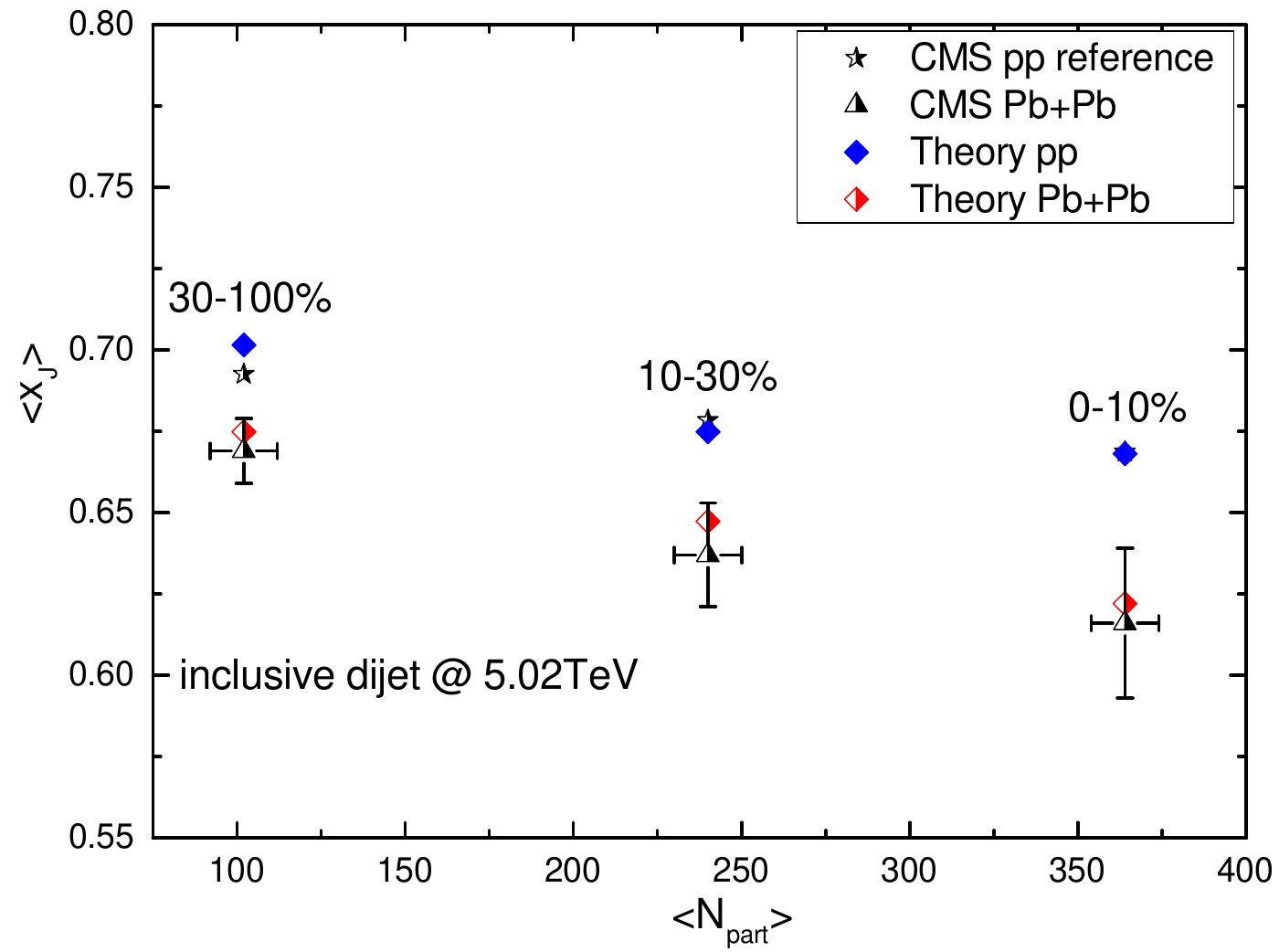}
\includegraphics[width=3.4in,height=2.8in,angle=0]{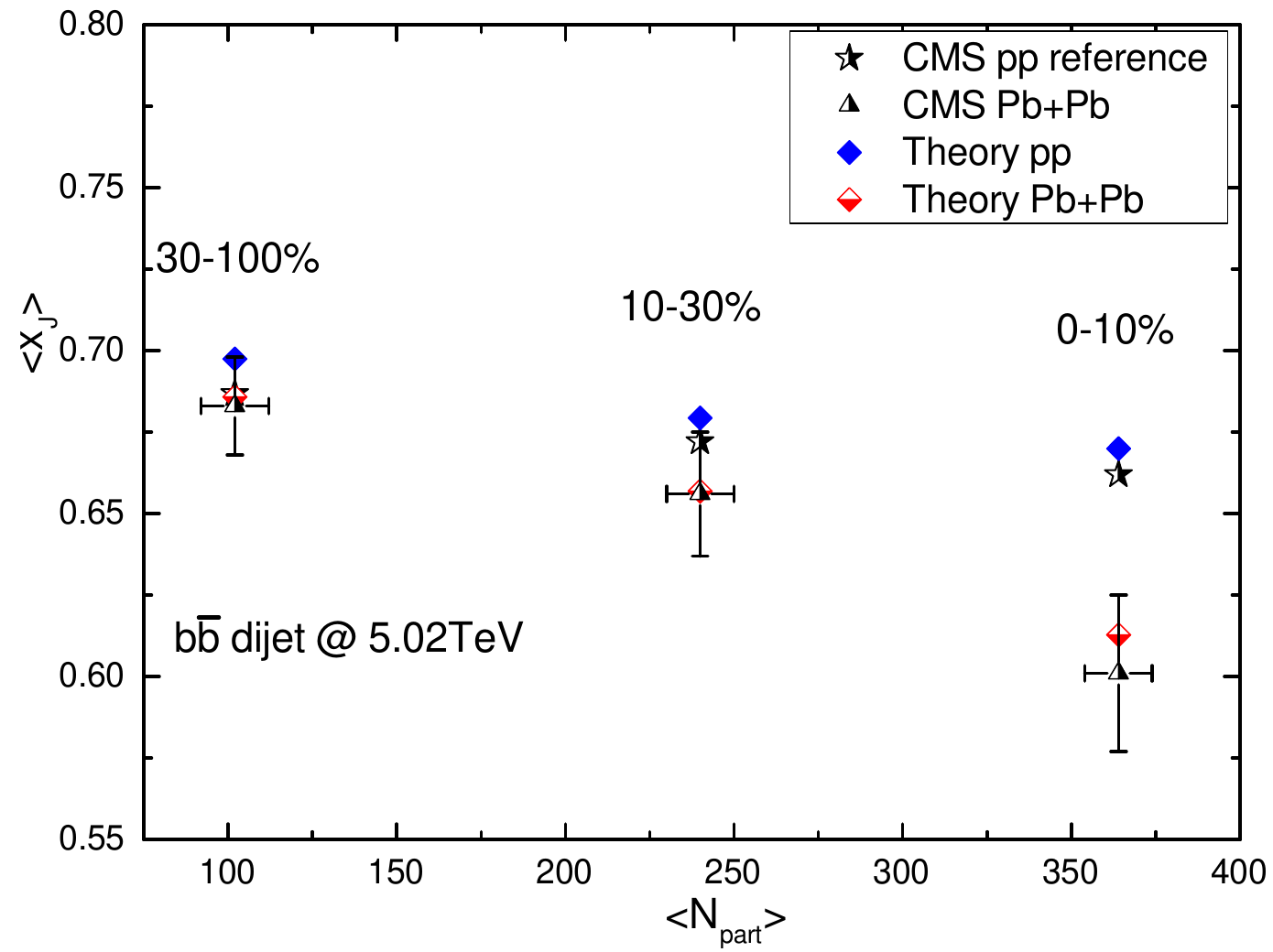}

\vspace*{-.2in}
\caption{Upper: Averaged $x_J$ in inclusive dijet production as a function of number of participant calculated in p+p and Pb+Pb collisions at different centralities compared with experimental p+p references and A+A data respectively; Bottom: Averaged $x_J$ in $b\bar{b}$ dijet production as a function of number of participant calculated in p+p and Pb+Pb collisions at different centralities compared with experimental p+p references and A+A data respectively.}
\label{fig:avexj}
\end{center}
\end{figure}

To further demonstrate the centrality dependence of the jet quenching effects on momentum balance of the $b\bar{b}$ dijets, we calculate the averaged $x_{\rm J}$ values as a function of the number of participants obtained in Pb+Pb collisions and the smeared p+p reference. The comparison of the values in systems with different centralities and the corresponding CMS data are shown in the bottom plots in Fig.~\ref{fig:avexj}. A good agreement between the theoretical calculations and experimental data is observed. We find the imbalance increases with the increasing centrality, even the averaged $x_{\rm J}$ of the p+p reference shift to smaller value with the increasing of centrality which is due to the resolution effects introduced by the experiment. More importantly, the averaged $x_{\rm J}$ value shift due to the jet quenching effect is much visible in central collision, but the imbalance in larger centrality such as $30-100\%$ Pb+Pb collision is compatible with their p+p reference which is unlike the case in dijets shown in the upper plots in Fig.~\ref{fig:avexj}, indicating a smaller energy loss than inclusive dijets in smaller centrality system.

\begin{figure}[!t]
\begin{center}
\vspace*{-0.2in}
\hspace*{-.1in}
\includegraphics[width=3.4in,height=2.6in,angle=0]{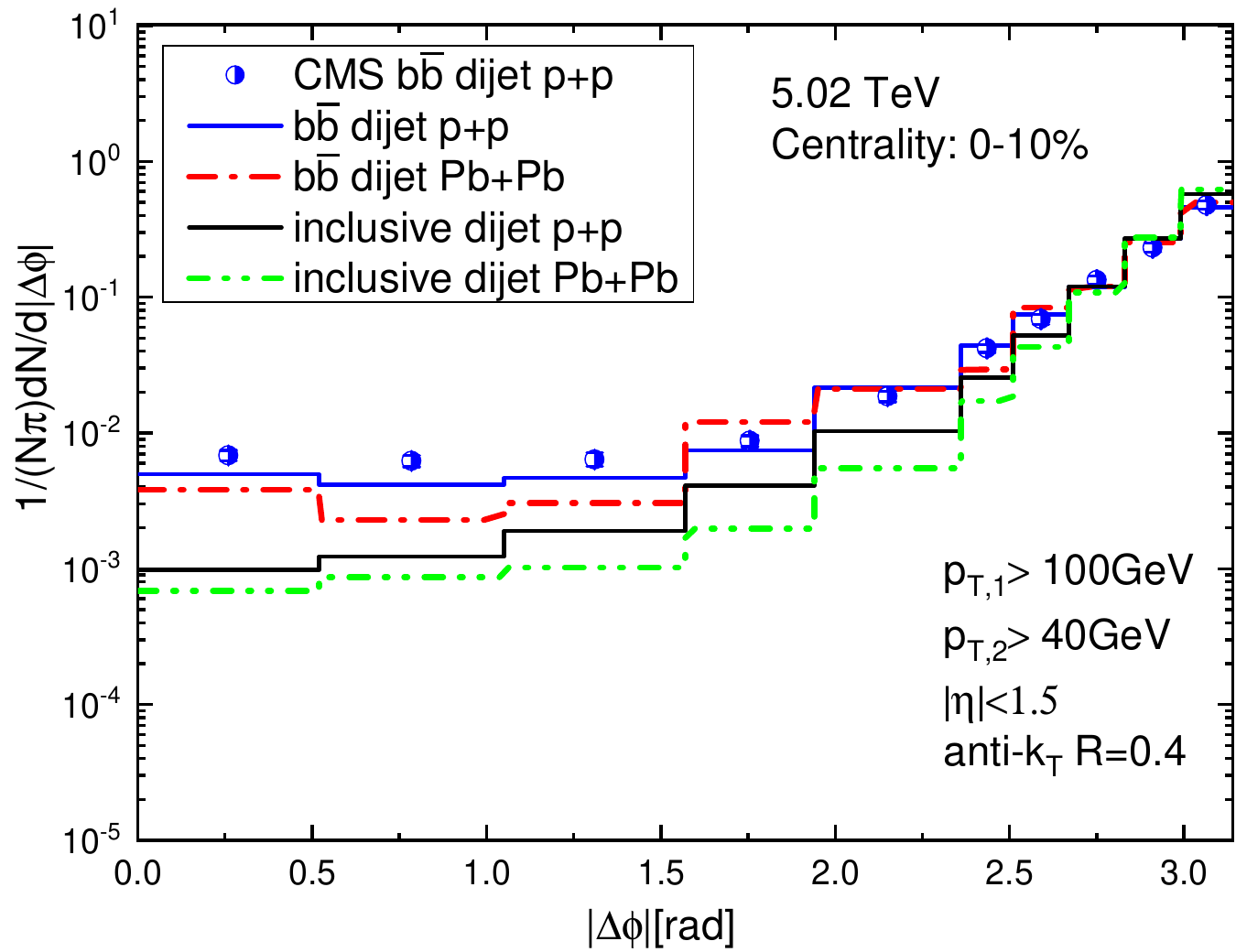}
\includegraphics[width=3.4in,height=2.6in,angle=0]{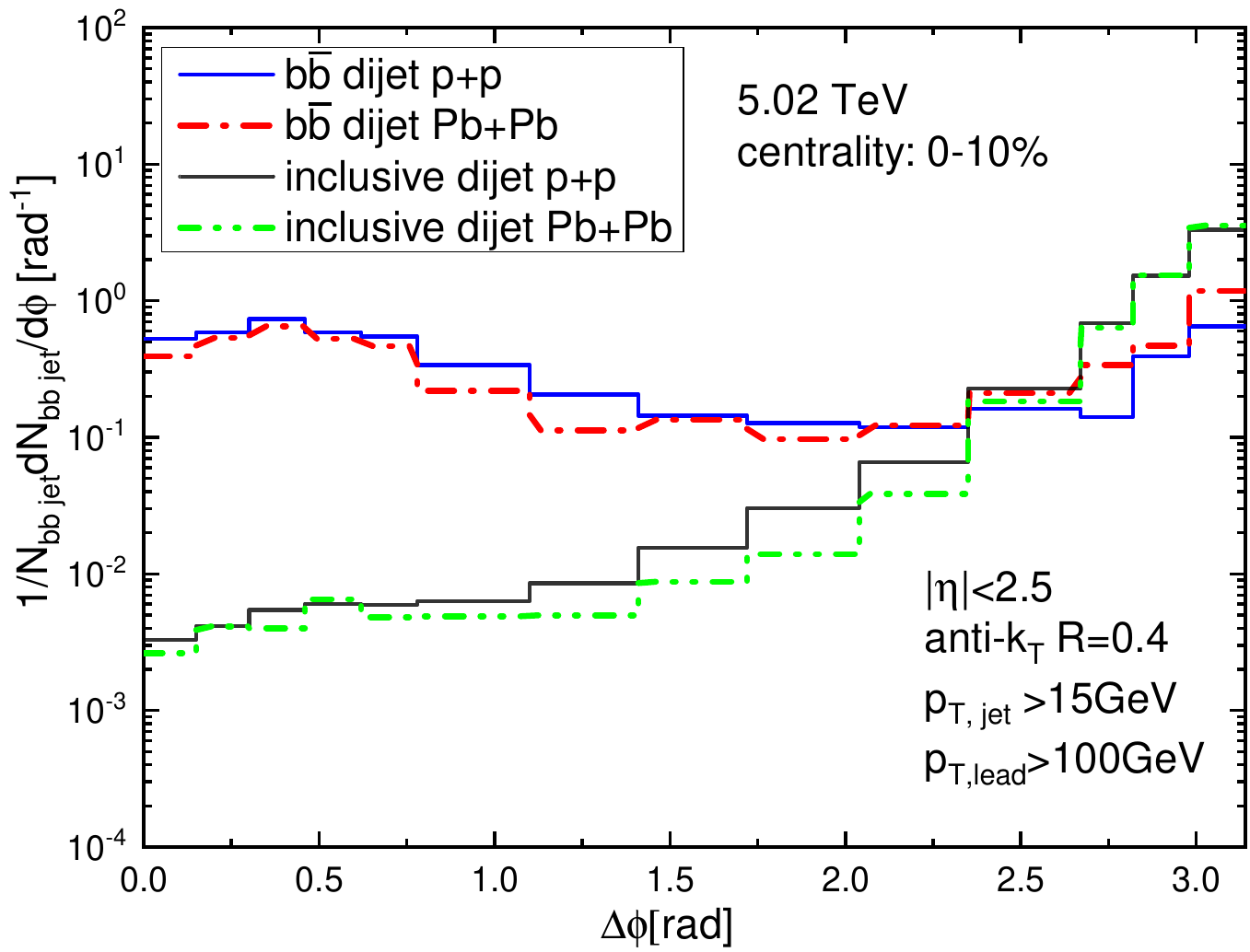}
\vspace*{-.2in}
\caption{Upper: the productions of $b\bar{b}$ dijets and inclusive dijets normalized by the number of events as functions of $\Delta \phi $ in p+p and Pb+Pb collisions at $5.02$~TeV TeV using the CMS configuration~\cite{Sirunyan:2018jju} comparing with the $b\bar{b}$ dijets p+p data; Bottom: the productions of $b\bar{b}$ dijets and inclusive dijets normalized by the number of events as functions of $\Delta \phi $ in p+p and Pb+Pb collisions at $5.02$~TeV using a lower minimum requirement of the jets $p_{\rm T}=15$~GeV. }
\label{fig:phi}
\end{center}
\end{figure}

Since the three production mechanisms can be experimentally separated by three event categories in azimuthal angle plane. It is essential to investigate the azimuthal angle distribution of the $b\bar{b}$ productions in p+p collisions and its modification in A+A collisions. we find its structure is quite sensitive to the jet event selection. When ATLAS define the dijet system as the minimum transverse momentum of the two highest-$p_T$ b-jets in an event should be $p_T > 20$~GeV and also $\vert \eta \vert < 2.5$~GeV, requiring their distance will be at least $\Delta R=0.4$ and the $p_T$ of the trigger jet should be larger than $270$~GeV~\cite{Aaboud:2016jed}, our simulation on the productions for $b\bar{b}$ dijets normalized by the number of events in p+p collision provided by SHERPA can describe the experimental data quite well (as seen in the upper plots of Fig.~\ref{fig:illustpp}),  especially at the same side peak ($\Delta \phi \to 0$)  which can be explained as the domination by the GSP process. Similar as the case in inclusive dijets, the angular correlation of $b\bar{b}$ dijet would also be modified by the hot and dense medium. We present the prediction of the medium modification for angular correlation of $b\bar{b}$ dijets and inclusive dijets in Pb+Pb collision at $\sqrt{s}=5.02$~TeV with centrality of $ 0-10\% $ using CMS configuration in the upper panel of Fig.~\ref{fig:phi}, we find the same side peak of the $b\bar{b}$ dijets disappear even in p+p collision comparing to the ATLAS measurement mentioned above, the energy loss effect will suppress the small $\Delta \phi$ distribution and enhance the distribution at large $\Delta \phi$ both for $b\bar{b}$ and inclusive dijets. In such selected b-jets, the b quark poses $80\% \to 95\%$ energy of the jets in p+p collision. But if we implement the configuration of ATLAS in Pb+Pb collisions at $5.02$~TeV, set the minimum transverse momentum of the two highest-$p_T$ b-jets (or jets) in an event should be $p_T > 15$~GeV and the leading jet $p_{\rm T}>100$~GeV in the bottom panel of Fig.~\ref{fig:phi}. We find the energy loss effect to the inclusive dijets production is similar as that in the upper plots and the energy loss effect to the $b\bar{b}$ dijets production would suppress and broaden the same side (small $\Delta \phi$) peak, also enhance and sharp the away side (near $\Delta \phi=\pi$ ) peak. But however, an overall suppression is found, it means, in the small angle region, it suffers a stronger suppression relative to the large angle region. Because that $b\bar{b}$ dijets produced at the small opening angle are dominated by GSP processes, and those b-jets are relatively ``softer'' than that produced by FCR. The energy loss effect would reduce the $p_{\rm T}$ of the low energy b jet to fall below the threshold of the event selection of $b\bar{b}$ dijets observable.

\section{Summary}
In summary, by using the NLO+PS event generator SHERPA, we implement a Monte Carlo simulation to take into account the collisional and radiative energy loss of the heavy and light quark in a hot and dense medium at the same time. For the first time, we present the theoretical calculation of the transverse momentum balance $x_J=p_{T,2}/p_{T,1}$ of the $b\bar{b}$ dijet in Pb+Pb collisions at $5.02$~TeV and confront them with the recent CMS measurement. We find that in $b\bar{b}$ dijets, similarly to that observed in inclusive dijets, the $x_{\rm J}$-distribution shift to smaller value due to the in-medium jet interaction. Furthermore, the comparison between the deviations of $\left\langle x_{J} \right\rangle_{pp}-\left\langle x_{J} \right\rangle_{PbPb}$ of inclusive dijet and $b\bar{b}$ dijets  at different centrality indicate smaller energy loss suffered in $b\bar{b}$ dijets than in inclusive dijets as the increasing of centralities.  At last, we find the energy loss effect will suppress the same side peak and enhance the away side peak of the normalized $\Delta \phi$ distributionin central Pb+Pb collisions at $5.02$ TeV, it is due to the fact that the near side peak distribution are dominated by the the contribution of GSP processes, and most of  them are low energy $b\bar{b}$ dijets, the $p_{\rm T}$ of the $b\bar{b}$ dijets can be easily reduced to fall below the kinetic cut.

{\bf Acknowledgments:}  The authors would like to thank D. Naplettano for helpful discussion about SHERPA, and  also P. Zhuang, K. Zhou Z. Xu and S. Chen for helpful discussions about heavy flavor transport. This research is supported by Natural Science Foundation of China with Project Nos. 11935007, 11805167.


\vspace*{-.6cm}

\begin{thebibliography}{99}

\bibitem{Wang:1991xy}
  X.~N.~Wang and M.~Gyulassy,
  Phys.\ Rev.\ Lett.\  {\bf 68}, 1480 (1992).

\bibitem{Gyulassy:2003mc}
  M.~Gyulassy, I.~Vitev, X.~N.~Wang and B.~W.~Zhang,
  In *Hwa, R.C. (ed.) et al.: Quark gluon plasma* 123-191
  [nucl-th/0302077].

\bibitem{Qin:2015srf}
  G.~Y.~Qin and X.~N.~Wang,
  Int.\ J.\ Mod.\ Phys.\ E {\bf 24}, no. 11, 1530014 (2015).


\bibitem{Vitev:2008rz}
  I.~Vitev, S.~Wicks and B.~W.~Zhang,
  JHEP {\bf 0811}, 093 (2008).

\bibitem{Vitev:2009rd}
  I.~Vitev and B.~W.~Zhang,
  Phys.\ Rev.\ Lett.\  {\bf 104}, 132001 (2010).



\bibitem{CasalderreySolana:2010eh}
  J.~Casalderrey-Solana, J.~G.~Milhano and U.~A.~Wiedemann,
  J.\ Phys.\ G {\bf 38}, 035006 (2011).



\bibitem{Zapp:2012ak}
  K.~C.~Zapp, F.~Krauss and U.~A.~Wiedemann,
  JHEP {\bf 1303}, 080 (2013).



\bibitem{Senzel:2013dta}
  F.~Senzel, O.~Fochler, J.~Uphoff, Z.~Xu and C.~Greiner,
  J.\ Phys.\ G {\bf 42}, no. 11, 115104 (2015).

\bibitem{Casalderrey-Solana:2014bpa}
  J.~Casalderrey-Solana, D.~C.~Gulhan, J.~G.~Milhano, D.~Pablos and K.~Rajagopal,
  JHEP {\bf 1410}, 019 (2014);
  Erratum: [JHEP {\bf 1509}, 175 (2015)].



\bibitem{Chang:2016gjp}
  N.~B.~Chang and G.~Y.~Qin,
  Phys.\ Rev.\ C {\bf 94}, no. 2, 024902 (2016).

\bibitem{Majumder:2014gda}
  A.~Majumder and J.~Putschke,
  Phys.\ Rev.\ C {\bf 93}, no. 5, 054909 (2016).



\bibitem{Chen:2016vem}
  L.~Chen, G.~Y.~Qin, S.~Y.~Wei, B.~W.~Xiao and H.~Z.~Zhang,
  Phys.\ Lett.\ B {\bf 773}, 672 (2017).

\bibitem{Chien:2016led}
  Y.~T.~Chien and I.~Vitev,
  Phys.\ Rev.\ Lett.\  {\bf 119}, no. 11, 112301 (2017).

\bibitem{Apolinario:2017qay}
  L.~Apolinário, J.~G.~Milhano, M.~Ploskon and X.~Zhang,
  arXiv:1710.07607 [hep-ph].

\bibitem{Connors:2017ptx}
  M.~Connors, C.~Nattrass, R.~Reed and S.~Salur,
  arXiv:1705.01974 [nucl-ex].

\bibitem{Dai:2012am}
  W.~Dai, I.~Vitev and B.~W.~Zhang,
  Phys.\ Rev.\ Lett.\  {\bf 110}, no. 14, 142001 (2013)
  [arXiv:1207.5177 [hep-ph]].

\bibitem{Wang:2013cia}
  X.~N.~Wang and Y.~Zhu,
  Phys.\ Rev.\ Lett.\  {\bf 111}, no. 6, 062301 (2013)
  [arXiv:1302.5874 [hep-ph]].

\bibitem{Zhang:2018urd}
  S.~L.~Zhang, T.~Luo, X.~N.~Wang and B.~W.~Zhang,
  arXiv:1804.11041 [nucl-th].

\bibitem{Neufeld:2010fj}
  R.~B.~Neufeld, I.~Vitev and B.-W.~Zhang,
  Phys.\ Rev.\ C {\bf 83}, 034902 (2011)
  [arXiv:1006.2389 [hep-ph]].


\bibitem{Qin:2010mn}
  G.~Y.~Qin and B.~Muller,
  Phys.\ Rev.\ Lett.\  {\bf 106}, 162302 (2011).

\bibitem{Young:2011qx}
  C.~Young, B.~Schenke, S.~Jeon and C.~Gale,
  Phys.\ Rev.\ C {\bf 84}, 024907 (2011).

\bibitem{He:2011pd}
  Y.~He, I.~Vitev and B.~W.~Zhang,
  Phys.\ Lett.\ B {\bf 713}, 224 (2012).

\bibitem{ColemanSmith:2012vr}
  C.~E.~Coleman-Smith and B.~Muller,
  Phys.\ Rev.\ C {\bf 86}, 054901 (2012).

\bibitem{Ma:2013pha}
  G.~L.~Ma,
  Phys.\ Rev.\ C {\bf 87}, no. 6, 064901 (2013).

\bibitem{Milhano:2015mng}
  J.~G.~Milhano and K.~C.~Zapp,
  Eur.\ Phys.\ J.\ C {\bf 76}, no. 5, 288 (2016).

\bibitem{Chen:2016cof}
  L.~Chen, G.~Y.~Qin, S.~Y.~Wei, B.~W.~Xiao and H.~Z.~Zhang,
  arXiv:1612.04202 [hep-ph].


\bibitem{Huang:2015mva}
  J.~Huang, Z.~B.~Kang, I.~Vitev and H.~Xing,
  Phys.\ Lett.\ B {\bf 750}, 287 (2015)
  [arXiv:1505.03517 [hep-ph]].


\bibitem{Huang:2013vaa}
  J.~Huang, Z.~B.~Kang and I.~Vitev,
  Phys.\ Lett.\ B {\bf 726}, 251 (2013)
  [arXiv:1306.0909 [hep-ph]].

\bibitem{Cao:2015hia}
  S.~Cao, G.~Y.~Qin and S.~A.~Bass,
  Phys.\ Rev.\ C {\bf 92}, no. 2, 024907 (2015)
  [arXiv:1505.01413 [nucl-th]].

\bibitem{Chen:2019gqo}
S.~Y.~Chen, B.~W.~Zhang and E.~K.~Wang,
Chin. Phys. C \textbf{44}, no.2, 024103 (2020)
doi:10.1088/1674-1137/44/2/024103
[arXiv:1908.01518 [nucl-th]].

\bibitem{Yan:2020zrz}
J.~Yan, S.~Y.~Chen, W.~Dai, B.~W.~Zhang and E.~Wang,
[arXiv:2005.01093 [hep-ph]].


\bibitem{Chen:2020pfa}
S.~Y.~Chen, W.~Dai, S.~L.~Zhang, Q.~Zhang and B.~W.~Zhang,
[arXiv:2005.02892 [hep-ph]].



\bibitem{Aaboud:2016jed}
  M.~Aaboud {\it et al.} [ATLAS Collaboration],
  Eur.\ Phys.\ J.\ C {\bf 76}, no. 12, 670 (2016)
  [arXiv:1607.08430 [hep-ex]].

\bibitem{Sirunyan:2018jju}
  A.~M.~Sirunyan {\it et al.} [CMS Collaboration],
  JHEP {\bf 1803}, 181 (2018)
  [arXiv:1802.00707 [hep-ex]].

  \bibitem{Norrbin:2000zc}
  E.~Norrbin and T.~Sjostrand,
  Eur.\ Phys.\ J.\ C {\bf 17} (2000) 137
  [hep-ph/0005110].

\bibitem{Combridge:1978kx}
  B.~L.~Combridge,
  Nucl.\ Phys.\ B {\bf 151} (1979) 429.

\bibitem{Nason:1987xz}
  P.~Nason, S.~Dawson and R.~K.~Ellis,
  Nucl.\ Phys.\ B {\bf 303} (1988) 607.

\bibitem{Beenakker:1990maa}
  W.~Beenakker, W.~L.~van Neerven, R.~Meng, G.~A.~Schuler and J.~Smith,
  Nucl.\ Phys.\ B {\bf 351} (1991) 507.

\bibitem{Banfi:2007gu}
  A.~Banfi, G.~P.~Salam and G.~Zanderighi,
  JHEP {\bf 0707}, 026 (2007)
  [arXiv:0704.2999 [hep-ph]].



\bibitem{Gleisberg:2008ta}
  T.~Gleisberg, S.~Hoeche, F.~Krauss, M.~Schonherr, S.~Schumann, F.~Siegert and J.~Winter,
  JHEP {\bf 0902}, 007 (2009)
  [arXiv:0811.4622 [hep-ph]].

\bibitem{Krauss:2001iv}
  F.~Krauss, R.~Kuhn and G.~Soff,
  JHEP {\bf 0202}, 044 (2002)
  [hep-ph/0109036].


\bibitem{Gleisberg:2008fv}
  T.~Gleisberg and S.~Hoeche,
  JHEP {\bf 0812}, 039 (2008)
  [arXiv:0808.3674 [hep-ph]].


\bibitem{Schumann:2007mg}
  S.~Schumann and F.~Krauss,
  JHEP {\bf 0803}, 038 (2008)
  [arXiv:0709.1027 [hep-ph]].

\bibitem{Nahrgang:2013saa}
M.~Nahrgang, J.~Aichelin, P.~B.~Gossiaux and K.~Werner,
  Phys.\ Rev.\ C {\bf 90}, no. 2, 024907 (2014)
  [arXiv:1305.3823 [hep-ph]].


\bibitem{Frixione:2002ik}
  S.~Frixione and B.~R.~Webber,
  JHEP {\bf 0206}, 029 (2002)
  [hep-ph/0204244].

\bibitem{Ball:2014uwa}
  R.~D.~Ball {\it et al.} [NNPDF Collaboration],
  JHEP {\bf 1504}, 040 (2015)
  [arXiv:1410.8849 [hep-ph]].



\bibitem{Cacciari:2011ma}
  M.~Cacciari, G.~P.~Salam and G.~Soyez,
  Eur.\ Phys.\ J.\ C {\bf 72} (2012) 1896
  [arXiv:1111.6097 [hep-ph]].

\bibitem{Khachatryan:2016jfl}
  V.~Khachatryan {\it et al.} [CMS Collaboration],
  Phys.\ Rev.\ C {\bf 96} (2017) no.1,  015202
  [arXiv:1609.05383 [nucl-ex]].

\bibitem{Jung:2014hja}
  K.~Jung [CMS Collaboration],
  Nucl.\ Phys.\ A {\bf 932} (2014) 253.


\bibitem{Moore:2004tg}
  G.~D.~Moore and D.~Teaney,
  Phys.\ Rev.\ C {\bf 71}, 064904 (2005)
  [hep-ph/0412346].


\bibitem{Rapp:2018qla}
  R.~Rapp {\it et al.},
  arXiv:1803.03824 [nucl-th].


  \bibitem{Scardina:2017ipo}
F.~Scardina, S.~K.~Das, V.~Minissale, S.~Plumari and V.~Greco,
  Phys.\ Rev.\ C {\bf 96}, no. 4, 044905 (2017)
  [arXiv:1707.05452 [nucl-th]].




 \bibitem{Djordjevic:2013xoa}
M.~Djordjevic and M.~Djordjevic,
  Phys.\ Lett.\ B {\bf 734}, 286 (2014)
  [arXiv:1307.4098 [hep-ph]].


\bibitem{Cao:2013ita}
S.~Cao, G.~Y.~Qin and S.~A.~Bass,
  Phys.\ Rev.\ C {\bf 88}, 044907 (2013)
  [arXiv:1308.0617 [nucl-th]].


\bibitem{Cao:2017hhk}
S.~Cao, T.~Luo, G.~Y.~Qin and X.~N.~Wang,
  Phys.\ Lett.\ B {\bf 777}, 255 (2018)
  [arXiv:1703.00822 [nucl-th]].





\bibitem{Beraudo:2015wsd}
 A.~Beraudo, A.~De Pace, M.~Monteno, M.~Nardi and F.~Prino,
  JHEP {\bf 1603}, 123 (2016)
  [arXiv:1512.05186 [hep-ph]].


\bibitem{Kang:2016ofv}
Z.~B.~Kang, F.~Ringer and I.~Vitev,
  JHEP {\bf 1703}, 146 (2017)
  [arXiv:1610.02043 [hep-ph]].




\bibitem{He:2011zx}
 M.~He, R.~J.~Fries and R.~Rapp,
  Phys.\ Rev.\ C {\bf 85}, 044911 (2012)
  [arXiv:1112.5894 [nucl-th]].


\bibitem{Lang:2012cx}
 T.~Lang, H.~van Hees, J.~Steinheimer, G.~Inghirami and M.~Bleicher,
  Phys.\ Rev.\ C {\bf 93}, no. 1, 014901 (2016)
  [arXiv:1211.6912 [hep-ph]].

\bibitem{Sharma:2009hn}
  R.~Sharma, I.~Vitev and B.~W.~Zhang,
  Phys.\ Rev.\ C {\bf 80}, 054902 (2009)
  [arXiv:0904.0032 [hep-ph]].

\bibitem{Zhou:2016vwq}
  K.~Zhou, W.~Dai, N.~Xu and P.~Zhuang,
  Nucl.\ Phys.\ A {\bf 956}, 120 (2016)
  [arXiv:1601.00278 [hep-ph]].

\bibitem{Wang:2019xey}
S.~Wang, W.~Dai, B.~W.~Zhang and E.~Wang,
Eur. Phys. J. C \textbf{79}, no.9, 789 (2019)
doi:10.1140/epjc/s10052-019-7312-4
[arXiv:1906.01499 [nucl-th]].

 \bibitem{Kubo}
  Kubo,
   Rep.\ Pro.\ Phys.{\bf 29} (1966) 255.

\bibitem{Gossiaux:2009qf}
  P.~B.~Gossiaux and J.~Aichelin,
  Nucl.\ Phys.\ A {\bf 830} (2009) 203C
  doi:10.1016/j.nuclphysa.2009.10.015
  [arXiv:0907.4329 [hep-ph]].

\bibitem{Plumari:2011mk}
  S.~Plumari, W.~M.~Alberico, V.~Greco and C.~Ratti,
  Phys.\ Rev.\ D {\bf 84} (2011) 094004
  doi:10.1103/PhysRevD.84.094004
  [arXiv:1103.5611 [hep-ph]].

\bibitem{Hosotani:2007kn}
  Y.~Hosotani, N.~Maru, K.~Takenaga and T.~Yamashita,
  Prog.\ Theor.\ Phys.\  {\bf 118} (2007) 1053
  doi:10.1143/PTP.118.1053
  [arXiv:0709.2844 [hep-ph]].

\bibitem{vanHees:2004gq}
  H.~van Hees and R.~Rapp,
  Phys.\ Rev.\ C {\bf 71} (2005) 034907
  doi:10.1103/PhysRevC.71.034907
  [nucl-th/0412015].

\bibitem{Francis:2015daa}
  A.~Francis, O.~Kaczmarek, M.~Laine, T.~Neuhaus and H.~Ohno,
  Phys.\ Rev.\ D {\bf 92}, no. 11, 116003 (2015)
  [arXiv:1508.04543 [hep-lat]].


\bibitem{Guo:2000nz}
  X.~f.~Guo and X.~N.~Wang,
  Phys.\ Rev.\ Lett.\  {\bf 85} (2000) 3591
  [hep-ph/0005044].

\bibitem{Zhang:2003wk}
  B.~W.~Zhang, E.~Wang and X.~N.~Wang,
  Phys.\ Rev.\ Lett.\  {\bf 93}, 072301 (2004)
  [nucl-th/0309040].

\bibitem{Zhang:2004qm}
  B.~W.~Zhang, E.~k.~Wang and X.~N.~Wang,
  Nucl.\ Phys.\ A {\bf 757} (2005) 493
  [hep-ph/0412060].




\bibitem{Majumder:2009ge}
  A.~Majumder,
  Phys.\ Rev.\ D {\bf 85} (2012) 014023
  [arXiv:0912.2987 [nucl-th]].

  \bibitem{Wang:2009qb}
  W.~t.~Deng and X.~N.~Wang,
  Phys.\ Rev.\ C {\bf 81} (2010) 024902
 [arXiv:0910.3403 [hep-ph]].

\bibitem{Chen:2010te}
  X.~F.~Chen, C.~Greiner, E.~Wang, X.~N.~Wang and Z.~Xu,
  Phys.\ Rev.\ C {\bf 81} (2010) 064908
  doi:10.1103/PhysRevC.81.064908
  [arXiv:1002.1165 [nucl-th]].

\bibitem{Shen:2014vra}
  C.~Shen, Z.~Qiu, H.~Song, J.~Bernhard, S.~Bass and U.~Heinz,
  Comput.\ Phys.\ Commun.\  {\bf 199} (2016) 61
  [arXiv:1409.8164 [nucl-th]].

\bibitem{Ma:2018swx}
  G.~Y.~Ma, W.~Dai, B.~W.~Zhang and E.~K.~Wang,
  Eur.\ Phys.\ J.\ C {\bf 79} (2019) no.6,  518
  [arXiv:1812.02033 [nucl-th]].

\bibitem{Cao:2018ews}
  S.~Cao {\it et al.},
  Phys.\ Rev.\ C {\bf 99} (2019) no.5,  054907
  [arXiv:1809.07894 [nucl-th]].

\bibitem{Burke:2013yra}
  K.~M.~Burke {\it et al.} [JET Collaboration],
  Phys.\ Rev.\ C {\bf 90} (2014) no.1,  014909
  [arXiv:1312.5003 [nucl-th]].


\end{thebibliography}
\end{document}